# Fast Microwave-free State Preparation and Measurement of Superconducting Qubits

Northrop Grumman Quantum Computing Team
*Northrop Grumman Mission Systems; Linthicum Heights, MD, 21090, USA*[*]

Fast, high-fidelity, scalable state preparation and measurement is critical to the realization of a quantum computing system. The state-of-the-art methods for preparation and readout of superconducting qubits require finely tuned microwave signals and ∼100 ns of measurement time, which are major obstacles to the scalability and performance of superconducting quantum computers. Here, we have demonstrated novel, microwave-free methods for both preparation and readout of superconducting qubits with >99% fidelity in only 10 ns for either operation while maintaining qubit coherence. This technology is compatible with scalable superconducting digital control systems, and using quantum flux parametrons for amplification, we demonstrated full quantum-to-digital conversion in only 15 ns, which is an order of magnitude faster than state-of-the-art microwave-based techniques.

## I. INTRODUCTION

Large scale quantum computing systems have the potential to unlock new advances in science and technology by performing previously intractable computations; however, actually building a useful quantum computer will require overcoming a vast array of challenges. A core requirement for scalable quantum computing is fast, accurate readout of the qubits. Superconducting qubit platforms often use a cQED dispersive readout technique [1–5] where even state-of-the-art devices require readout times 10-100x longer than their gate times. These long readout times substantially increase the runtime of quantum algorithms [6]. Furthermore, high-fidelity single-shot dispersive readout requires quantum-limited low-noise amplifiers that have low saturation powers, inhibiting the scaling of the technology to many qubits [7, 8].

Another immediate obstacle for achieving a large-scale system is reducing the required amount of room temperature control lines and hardware. Although cryoCMOS control [9] provides a partial solution, interfacing the cryoCMOS controller with the superconducting qubit chip is a monumental thermal and electrical engineering challenge. Superconducting single-flux-quantum (SFQ) digital logic provides a low-power, low-temperature command and control solution that can coexist in the same package with the quantum processor, thus eliminating all direct-control cabling [10]. However, the SFQ driven technologies demonstrated to date require a microwave tone to be delivered to the millikelvin stage in order to prepare the qubits and perform gates [11], which is a major deterrent to scalability.

Here, we present a novel state preparation and measurement (SPAM) architecture that is fast, fully compatible with SFQ-based (e.g. ERSFQ, RQL, AQFP, etc.) control systems, and eliminates the need for microwave tone coupling to quantum circuits. We demonstrated this solution with consistent fidelities approaching 99% and an optimal readout operation time of approximately 10 ns as measured across 282 qubits, with top-performing devices demonstrating 99.7% fidelity. These devices also achieved a full quantum-to-digital transformation in 15 ns through the use of quantum flux parametrons (QFPs) for isolation and amplification [12] and an analog-to-digital converter (ADC) [13].

We achieved this fast, single-shot readout by strongly coupling a flux qubit (FQ) to a high-coherence computational qubit (CQ). The strong coupling enabled a high-fidelity adiabatic swap of the photonic state of the CQ into the FQ. Then we completed the FQ anneal, which adiabatically converted the photonic state into a long-lived classical circulating current state. We performed the full readout in a single annealing operation with a linear flux ramp, which is compatible with SFQ-driven on-chip digital-to-analog converters (DACs).

Using only ramped flux signals, we demonstrated a novel qubit preparation method by way of rapid adiabatic passage (RAP) [14]. We manipulated the FQ energy potential to create a single-photon excitation in the system and swapped it into the CQ. Preparation of the $|1\rangle$ state was essentially the reciprocal operation to readout in this scheme and thus could be performed in only 10 ns. Furthermore, we demonstrated the preparation of coherent superposition states using our RAP preparation technique. By analyzing the Ramsey oscillations, we measured the dephasing rate of the CQ and observed its spectrum without using microwaves to excite it.

## II. COMPUTATIONAL QUBIT-FLUX QUBIT SYSTEM

### A. Measurement

In order to perform complex quantum operations, such as quantum error correction, it is necessary to be able to quickly measure and convert the qubit's quantum state into a classical signal that can interface with control hardware. Flux qubits provide a natural option to perform this conversion, as the flux qubit eigenstates have DC cir-culating currents under the right bias conditions. In our previous work we have demonstrated an amplification of FQ circulating current states through a series of quantum flux parametrons (QFPs) [12] to the level necessary to trigger digital logic [13]. However, to quickly generate circulating currents in the FQ its capacitance must be kept low, which is known to be detrimental to qubit coherence. We, instead, leveraged a computational qubit (CQ), which had a highly-coherent resonant mode that was prepared and read out by the FQ. This mode was positioned far in frequency space from the upper sweet spot of the FQ, thereby mitigating any loss from the hybridization of the modes during gate operations. By carefully engineering this CQ-FQ system, we enabled high fidelity (>99%) readout in 10 ns while maintaining the coherence time of the computational state.

[*] See acknowledgements for author list.

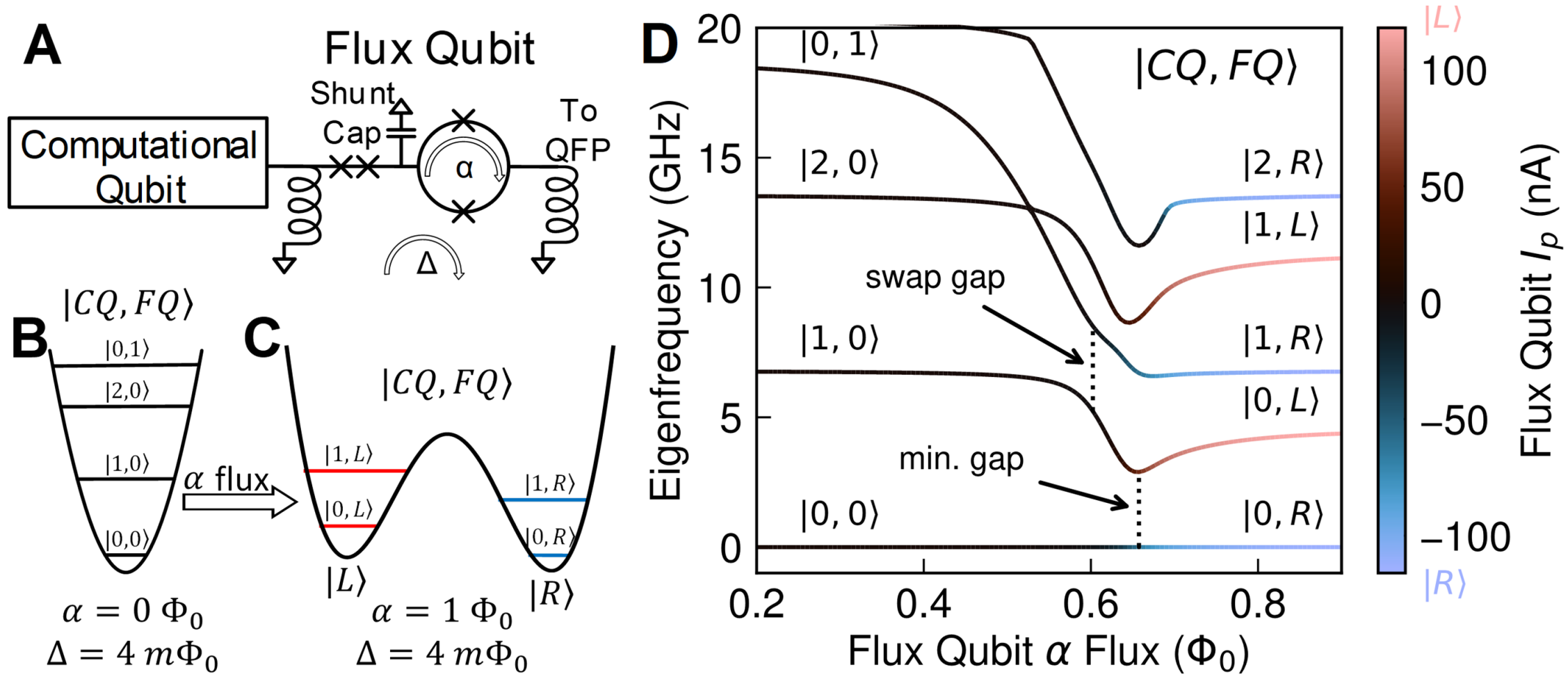

FIG. 1. Converting a qubit state to a circulating current. (A) The flux qubit (FQ) used for readout with an inductive coupling to the computational qubit (CQ). (B) Illustration of the FQ potential with plasmon-like states at $\alpha = 0$ (C) Double-well potential of the FQ at $\alpha = \Phi_0$ and $\Delta = 4\ m\Phi_0$. The states in different wells have opposite circulating currents. The $\Delta$ flux breaks the degeneracy between the wells. (D) Model device eigenspectrum as a function of annealing flux, $\alpha$, with $\Delta = 4\ m\Phi_0$. States are labeled in the $|CQ, FQ\rangle$ basis at the beginning and end of the readout process. The first excited state maintained a large energy separation from other levels the annealing process, which allowed the $|1, 0\rangle$ state to adiabatically transform from a photonic excitation to a circulating current state $|0, L\rangle$. The color of the levels denotes the circulating current in the FQ, $I_p$. The excitation in the CQ, $|1, 0\rangle$, swaps into the FQ, $|0, 1\rangle$, at the swap gap and is converted to a circulating current state, $|0, L\rangle$ at the min. gap. These gaps must be large to avoid errors from Landau-Zener transitions. The measured spectrum was used to validate the circuit Hamiltonian model in Fig. S1 of the supplemental information.

The CQ photonic mode may be coupled to the FQ through either an inductive (shown in Fig. 1A) or capacitive coupling. Fluxes applied to two loops, labeled $\alpha$ and $\Delta$, were used to control the FQ. An applied $\alpha$ flux transitioned the FQ from a single-well potential with a high-frequency first excited state to a double-well potential (from B to C in Fig. 1, referred to as ‘annealing’). The states in the double-well had persistent circulating current in the $\Delta$ loop flowing in opposite directions, $|L\rangle$ and $|R\rangle$. We adjusted the tilt of the potential by applying a $\Delta$ flux, which made the $|R\rangle$ well lower in energy than the $|L\rangle$ well (Fig. 1C).

We read out the CQ state in one fluid step that consisted of two key quantum processes: the CQ/FQ swap and the current state projection in the FQ. The readout may be understood through the simulated spectrum shown in Fig. 1D. The CQ/FQ system initially idled on the left side of the plot at low $\alpha$ flux where the CQ could perform quantum gates without being affected by the FQ. At that point, both the CQ (∼7 GHz) and FQ (∼18 GHz) were fully photonic and no persistent current existed in the FQ $\Delta$ loop. As we increased the $\alpha$ flux for readout, the FQ frequency came down and swept through an engineered swap gap with the CQ at about 0.6 $\Phi_0$. As the system passed through the large (∼1-2 GHz) anti-crossing, the nature of the excited state changed adiabatically from being an excitation in the CQ to being in the FQ. We expected this adiabatic swap to be successful if the sweep rate was slow compared with the swap gap [15–17]. Next, we completed the anneal of the flux qubit, which projected the photonic state into a highly stable and essentially classical circulating current state in a double-well potential [12, 18]. This quantum-to-classical conversion began at the minimum energy gap (min. gap) labeled in Fig. 1D, after which persistent current grew rapidly as indicated in the plot by the color of the spectral lines. Similarly, the ground state of the CQ mapped to the lower energy well of the flux qubit and thus the opposite circulating current state. Simply by ramping the $\alpha$ flux on the FQ, we converted the $|0\rangle$ and $|1\rangle$ states of the CQ into macroscopic circulating currents (∼100 nA), which could readily be sensed by digital SFQ electronics such as an adiabatic quantum flux parametron (AQFP) [19–21], SFQ comparator [22], or RQL analog-to-digital converter [13].

In order to faithfully project the $|0\rangle$ and $|1\rangle$ states of the CQ onto the $|L\rangle$ and $|R\rangle$ circulating current states of the FQ, we had to apply a particular $\Delta$ flux offset to the FQ during the annealing operation. Although the min. gap of the spectrum was expected to grow with increasing $\Delta$, we approached a practical limit on that bias as the energy separation between $|1\rangle$ and $|2\rangle$ began to close with increasing $\Delta$. At the point where $|1\rangle$ and $|2\rangle$ were degenerate in the double-well regime, we expected to see macroscopic resonant tunneling (MRT) [23] between the circulating current states, which caused errors. We refer to the range of $\Delta$ control where we saw good $|0\rangle$ and $|1\rangle$ visibility as the 1st MRT. Outside of the 1st MRT, the readout gave the same result for both $|1\rangle$ and $|0\rangle$ because both states adiabatically evolved into states with circulating current flowing in the same direction. We found that this region could easily be designed to be $\sim 10\ m\Phi_0$ (see Fig. 2B) making the readout exceptionally robust against flux drift. It was also insensitive to the frequency of the CQ. We designed the FQ such that its upper sweet spot (USS) in $\alpha$ flux was substantially higher than the CQ frequency, providing wide control margins with respect to the start and stop points of the flux qubit $\alpha$ annealing ramp.

To perform high fidelity adiabatic readout of the CQ with this method, there were two competing time scales. If the flux qubit was annealed too quickly, non-adiabatic effects would dominate, causing Landau-Zener-like transition errors at both the swap gap and the min. gap [16, 17]. If annealed too slowly the excitation in the system would decay before the barrier was fully raised between the circulating current wells. In order to optimize our readout performance, we swept the FQ $\alpha$ rise time as seen in Fig. 2C. At short rise times (<5 ns) we saw that the readout fidelities of the $|0\rangle$ and $|1\rangle$ states were poor and dominated by Landau-Zener transition errors. We observed an exponential drop in readout error as a function of rise time with a characteristic time constant ($\tau_{LZ}$). Generally speaking, $\tau_{LZ}$ could be decreased by increasing the swap gap (allowing for a faster adiabatic swap) and increasing or smoothing out the minimum gap to minimize non-adiabatic scattering. We used time-dependent simulations of the readout process to guide circuit parameter selection and found that there was a complex relationship between capacitance, critical current, and inductance in the FQ with respect to readout fidelity. For the devices described in this paper, we found $\tau_{LZ}$ to be approximately 1.5 ns.

At long rise times (>100 ns) the readout fidelity became limited by decay during the readout process. The $|1\rangle$ state readout fidelity dropped exponentially with a timescale, $\tau_D$, and we found this was due to a combination of effects. First, the FQ lifetime set a background decay rate that increased just after the swap as the circulating current began to develop [24] and before the barrier between circulating current wells was fully in place. Second, the large frequency excursion made by the $|1\rangle$ state during the readout process resulted in energy exchange between the FQ and any strongly-coupled two-level systems (TLSs), governed by the Landau-Zener transition probability for each TLS. Ramping faster increased the energy velocity, and therefore reduced the probability of transitioning the state from the flux qubit to the TLS. Additional discussion of this effect can be found in supplemental material. At very long rise times (>10 $\mu$s) the readout error increased due to thermalization with the environment at the relatively low min. gap frequency. We optimized the readout by balancing the benefits of annealing faster to avoid energy dissipation and TLS interactions against the errors incurred from annealing too fast, such as lack of adiabaticity and scattering into nearby states.

### B. State Preparation

In order to prepare the CQ in the $|1\rangle$ state, we essentially ran the adiabatic readout process in reverse with a slight modification to $\Delta$ flux. We illustrate the Rapid Adiabatic Passage state preparation (RAP prep) process in Fig. 2A where we traversed the $\alpha$-$\Delta$ flux control space in such a way as to create an excitation in the FQ and then adiabatically swapped it into teh CQ. We began the process with the FQ at low $\alpha$ with a single-well potential and in the ground state (1). Just as in the readout process, we annealed the FQ to project the ground state into the $|L\rangle$ well (1→2). We then applied a $\Delta$ flux pulse that tilted the double-well potential, thus making $|L\rangle$ the higher energy state (2→3). Then we unannealed the FQ, creating a photonic excitation in the FQ and swapping it into the CQ (3→4). Once the excitation was in the CQ and the FQ was at low $\alpha$ bias, we removed the $\Delta$ flux pulse (4→1). We demonstrated in Fig. 2B our ability to prepare the CQ in the $|1\rangle$ and read it out with high-fidelity. The dashed lines in Fig. 2B came from a simulation of the Hamiltonian in Eq. S1, where the only free parameter was the time between RAP Prep and measure for the Prep $|1\rangle$ data and was set to match experimental timing. The target for the $\Delta$ pulse amplitude (2→3) was $\sim 10\ m\Phi_0$ wide, similar to the margin for the $\Delta$ offset in the readout operation. In contrast to microwave-based qubit state preparation methods (which must be carefully calibrated with respect to drive frequency and phase, pulse envelope, DRAG parameters, etc.), RAP prep was entirely agnostic to the frequency of the CQ and had wide control margins for all pulse parameters including offset, height, and ramp rate.

RAP preparation fidelity was limited by the same factors that limited readout, $\tau_{LZ}$ and $\tau_D$, and the optimal preparation time was similarly 10 ns (Fig. 2C). We have compiled fidelity measurements from 282 devices across six different fabrication lots (Fig. 2D). The SPAM rates we report are the combined error from both preparation and measure operations. We observed higher SPAM error for prepare $|1\rangle$ commands primarily due to photon loss ($\tau_D$) during the flux qubit anneal and additional Landau-Zener errors incurred at the swap gap ($\tau_{LZ}$). We also used RAP prep to prepare and explore the behavior of higher excited states (i.e. $|2\rangle$, $|3\rangle$, etc.) by adjusting the $\Delta$ flux offset and pulse height to access different MRTs as discussed in the supplemental information (see Fig. S2).

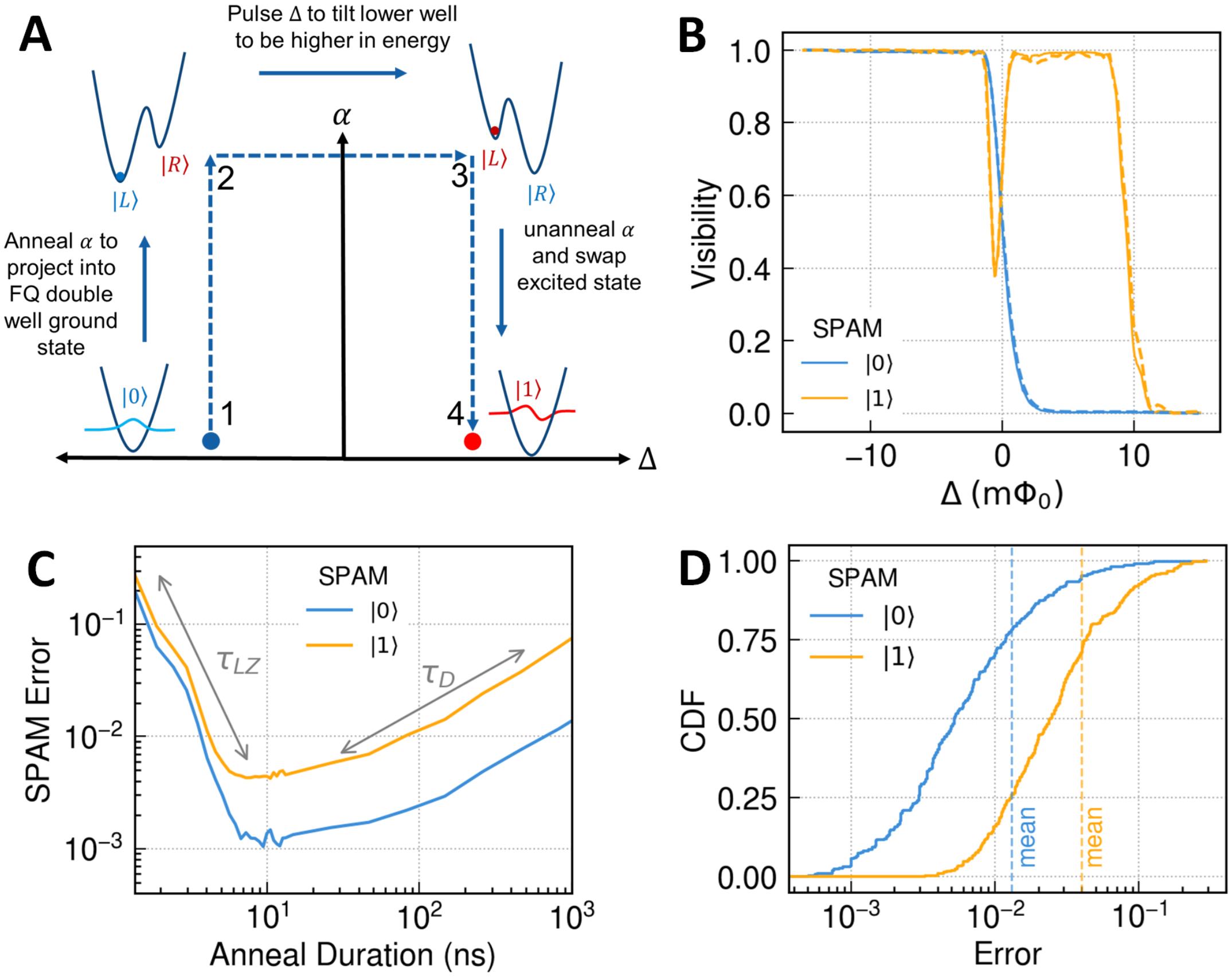

FIG. 2. High-fidelity state preparation and measurement using a flux qubit. (A) Rapid adiabatic passage preparation (RAP prep) of an excited state in the computational qubit (CQ) using a flux qubit (FQ). Beginning in the ground state (1), the FQ was annealed to project the state into the $|L\rangle$ well (1→2). Then the double-well potential was tilted until the prepared well became the excited state (2→3). The FQ was then unannealed, transitioning back to a single-well potential and swapping the photon into the CQ (3→4). (B) FQ readout calibration plot showing high-fidelity state preparation and measurement (SPAM) across a robust calibration window. The dashed lines are time-dependent simulations of RAP prep on the circuit Hamiltonian model that matched the measured spectrum in Fig. S1B. (C) Measured SPAM error as a function of FQ $\alpha$ anneal duration. For short anneal durations (<5 ns) the readout was not adiabatic and the error was dominated by Landau-Zener ($\tau_{LZ}$) transitions. At long anneal durations (>100 ns) the error became dominated by decay of the excited state ($\tau_D$). The optimal anneal duration was approximately 10 ns for both preparation and readout. (D) Distribution of SPAM errors from N=282 devices across six different fabrication lots.

In addition to preparing high-fidelity $|0\rangle$ and $|1\rangle$ states, we demonstrated the direct preparation of coherent superpositions of $|0\rangle$ and $|1\rangle$ in the CQ with RAP prep. By shifting the $\Delta$ offset to the degeneracy point during the unanneal ($\Delta$=0 from 3→4, see Fig. 3A), we prepared the qubit in a coherent 50/50 superposition state. This directly prepared the CQ on the X axis of the Bloch sphere (i.e. $|X\rangle = \frac{1}{\sqrt{2}}(|0\rangle + |1\rangle)$ in only 10 ns. Because this operation time was short compared to the timescale associated with any residual splitting, the circulating current state evolved under the sudden approximation into a superposition of ground and excited states. Likewise, we measured the CQ in the $|X\rangle$ basis directly by annealing the FQ while it was biased at degeneracy.

Fig. 3B shows the results of a Ramsey experiment where we prepared the CQ along $X$, delayed for some time and then read out along $X$. We observed the expected Ramsey oscillation with a decay time, $T_2^* = 24.0(6)$ $\mu$s, and found that the frequency of the oscillation matched the $f_{01}$ of the CQ that we measured with microwave spectroscopy (see supplemental information Fig. S1B) when we accounted for the timing resolution of our control electronics. We also confirmed that the CQ frequency varied with $\alpha$ idle flux as expected (Fig. 3C). Note that the CQ maintained a relatively long coherence time despite being strongly coupled to the FQ because of the large FQ/CQ detuning at idle.

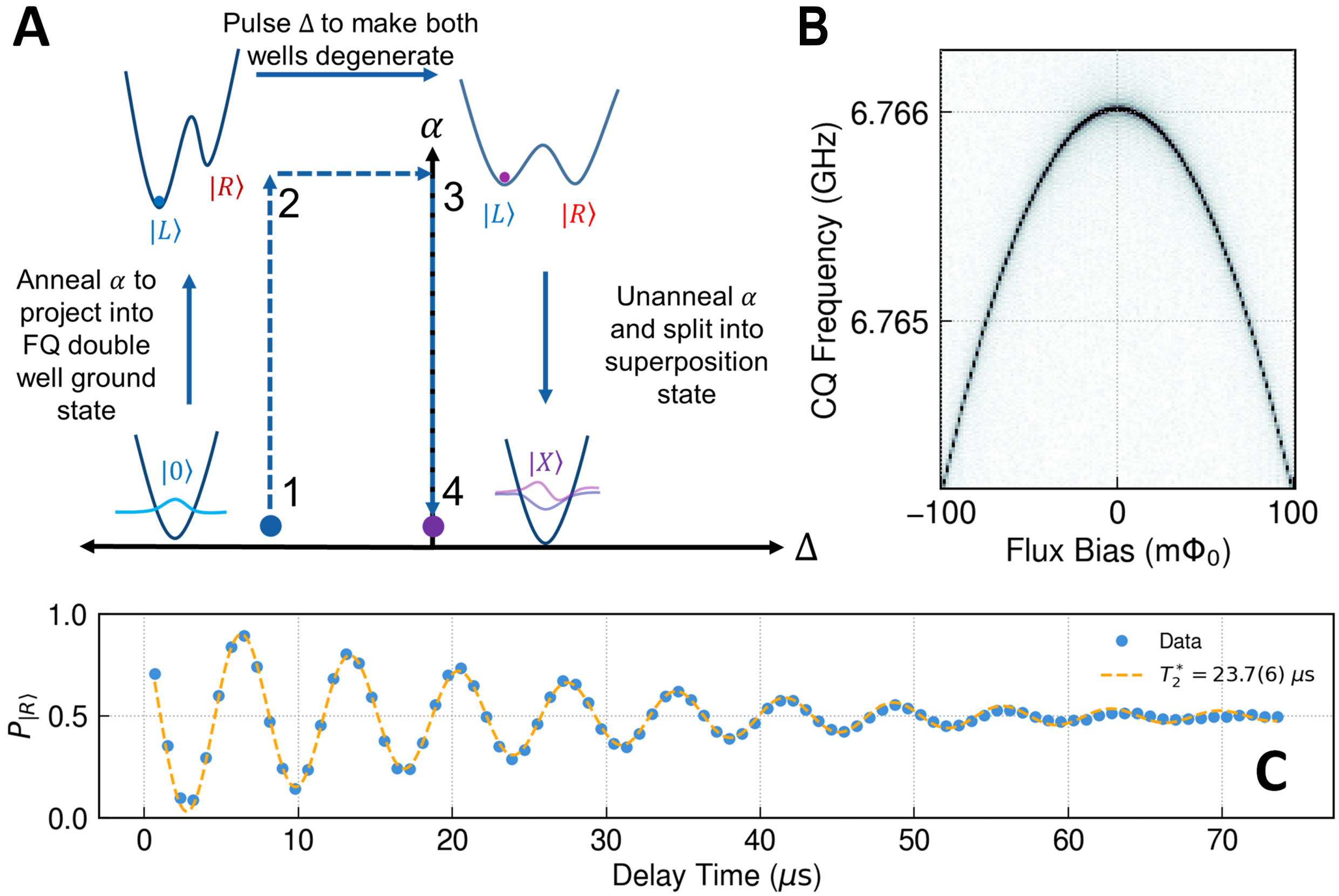

FIG. 3. Superposition State Preparation. (A) RAP preparation of a computational qubit (CQ) in a quantum superposition state using a coupled flux qubit (FQ). First the FQ was annealed to project the ground state to the $|L\rangle$ well (1→2). Then the double-well potential was tilted until the two wells were degenerate (2→3). The FQ was then unannealed, transitioning back to a single-well potential and splitting the state into a superposition of $|0\rangle$ and $|1\rangle$ in the CQ (3→4). (B) Microwave-free spectroscopy of the CQ frequency as a function of a tunable flux bias. Shown here as the FFT of a Ramsey signal acquired using the protocol depicted in A. (C) Coherent Ramsey oscillation of a superposition state prepared with the same protocol. For both B and C, the CQ was also measured in the $|X\rangle$ basis by annealing the FQ at degeneracy (4→3). Our preparation and measurement technique enables microwave-free spectroscopy of the CQ and rapid characterization of qubit properties (e.g. $T_2^*, T_1$).

This preparation of the CQ to a superposition state has significant advantages for tuning and characterizing qubits in the lab. For example, using RAP prep $|X\rangle$ we could perform microwave-free spectroscopy on the quantum system as in Fig. 3B. Unlike pulsed microwave gates and dispersive readout schemes, the preparation and readout methods we present in this paper worked independently of the qubit frequency (so long as the CQ frequency was between the FQ min. gap and USS). This made it quick and easy to investigate qubit properties, such as lifetime and dephasing, as a function of various qubit biases.

## III. QUANTUM-TO-DIGITAL TRANSFORMATION

Once we converted the quantum information in the CQ into classical circulating current in the FQ, the next step was to feed that information into the digital SFQ data stream. As we showed in earlier publications [12], we use quantum flux parametrons (QFPs) as our link to the SFQ logic (Fig. 4A) because they provide two critical advantages: they isolate the qubit from the noise and resistors in the digital circuitry and they amplify the FQ signal to a level that is compatible with SFQ logic. Here we showed that the QFPs can be annealed much more quickly than the FQ while providing the amplification needed for analog-to-digital conversion in only 5 additional nanoseconds while contributing minimal (∼1E-5) additional error (Fig. 4B).

We used the QFP as a very sensitive flux sensor that provided isolation during quantum operations [12] and swift amplification during readout. By tuning the QFP $\Delta$ flux to 0 and galvanically coupling to the FQ, we positioned the QFP such that the current in the FQ tilted the QFP potential and in turn induced a right or left circulating current in the annealed QFP based on the state of the FQ. Each QFP stage reliably provided an order of magnitude amplification of the circulating current. Like the FQ, the speed limit on the QFP anneal was set by the characteristic Landau-Zener time when passing through the min. gap, and the min. gap was proportional to the $\Delta_{QFP} I_c$ where $\Delta_{QFP}$ is the flux offset provided by the FQ circulating current and $I_c$ is the critical current of the QFP junctions. Therefore, we designed our QFPs with much larger critical currents than the FQ and coupled them to FQ such that $\Delta_{QFP} \sim 10\ m\Phi_0$. Note that, in contrast to the FQ, which was limited to $\Delta$ biases within the 1st MRT during readout so as to distinguish $|0\rangle$ and $|1\rangle$ population, the QFP was always meant to remain in the ground state, so we could design the coupling between the FQ and QFP to provide maximal signal, limited mainly by layout footprint considerations. This increased the min. gap frequency and enabled much faster QFP annealing times without sacrificing adiabaticity.

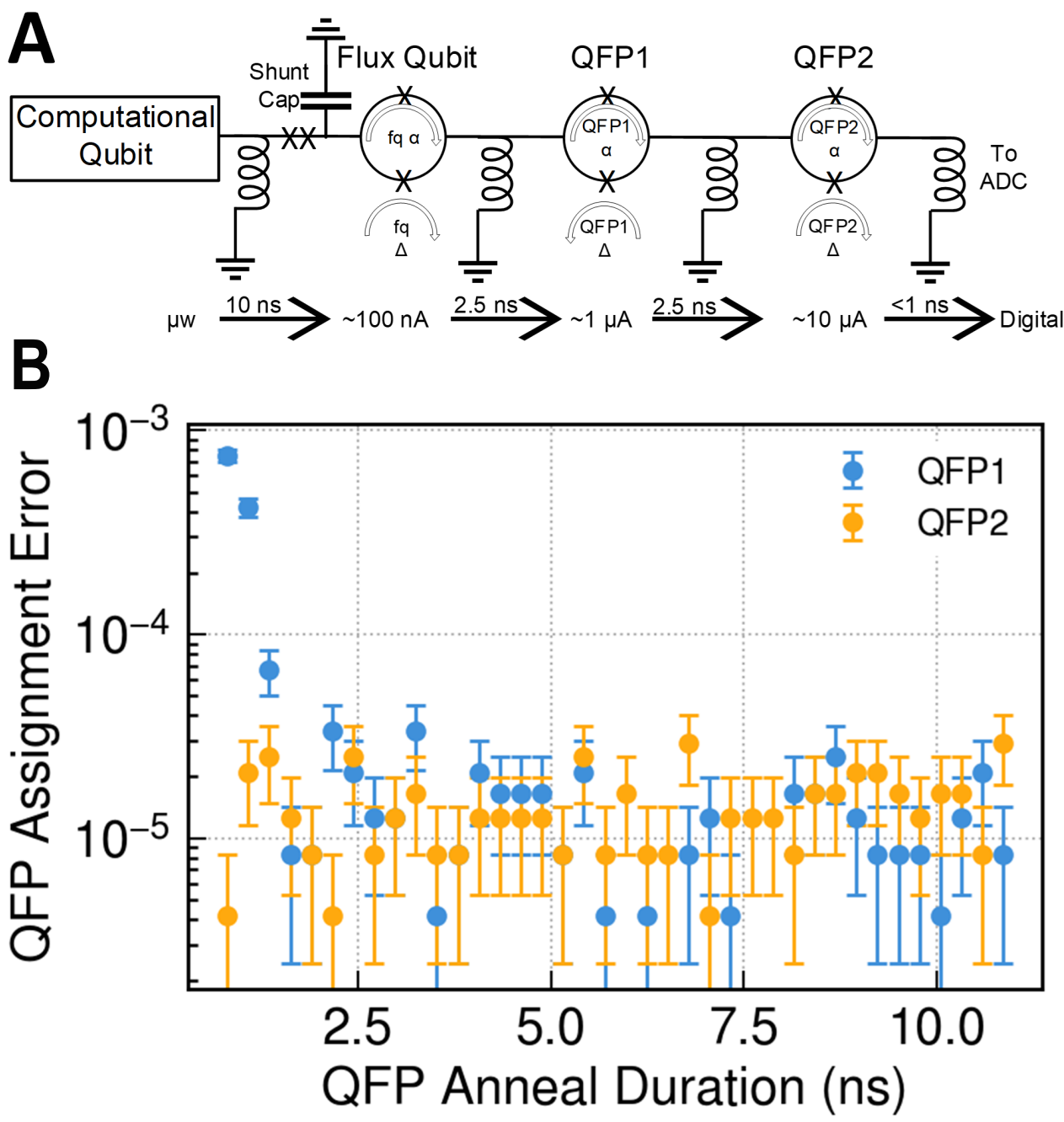

FIG. 4. Signal Amplification for On-Chip Logic. (A) Transitioning the quantum state of the computational qubit (CQ) into a DC circulating current in the flux qubit (FQ) produced circulating currents of order 100 nA. In order to amplify this current until it was large enough to be converted to a digital signal, a series of QFPs were used, each with increasing critical current. The QFPs could then be coupled to an Analog-to-Digital Converter (ADC) (in the fashion shown here), or to a flux tunable circuit element such as a tunable resonator. (B) State assignment error of the two QFP chain as a function of the QFP anneal time. Both QFPs could be annealed in <2.5 ns while maintaining low readout error (of order 1E-5). Error bars come from shot noise uncertainty on the measurement.

Fig. 4 shows representative data of the readout error of a two-QFP chain. QFP1 was coupled to the FQ and had junctions with critical currents an order of magnitude larger than the FQ. We achieved an error rate as low as 1E-5 in QFP1 for anneal times >2 ns. Since the critical currents of the junctions in QFP2 were an order of magnitude larger than QFP1, we were not able to detect any significant increase in the error in QFP2 at even the fastest anneal speeds available in the test setup. So as not to convolve the quantum SPAM with the readout fidelity of the QFP chain, in this experiment we left the FQ unannealed and applied a $\Delta$ flux offset to QFP1 that was equivalent to the signal that would be provided by the fully-annealed FQ as the polarization bias. The ratio of this bias to the natural width of the $\Delta$ transition of QFP1 set the error floor of 1E-5 [12]. Further optimization of the FQ/QFP coupling is expected to reduce this error floor in future experiments.

Once we amplified the FQ current signal with the QFPs we coupled it into either an Analog-to-Digital Converter (ADC) [13] to convert it into a digital signal at the rate of the digital circuit's clock (~GHz) [25], or a flux-tunable element such as a tunable resonator [12] to measure it using standard microwave techniques. The former was much faster, fully converting the quantum state of the qubit into digital information in just 15 ns. This rapid quantum-to-digital conversion represents a scalable readout solution that could be integrated into a larger superconducting logic circuit. The latter was experimentally simpler and provided a means to explore the flux qubit readout technique independent of digital control. We successfully performed readout with both methods and used both in the development of this article.

## IV. DISCUSSION

Since superconducting qubit gate times are short compared to state-of-the-art measurement times, we require a new innovation in readout that is fast, accurate and compatible with a scalable control and measurement architecture. We have demonstrated state preparation and measurement of the $|0\rangle$ and $|1\rangle$ states of a qubit with up to 99.7% fidelity using only ramped flux signals and without the need for shaped microwave $\pi$ pulses. In a single operation with wide control margins, we adiabatically swapped the excited state population of a computational qubit (CQ) into a flux qubit (FQ) and then converted the photonic eigenstates of the FQ into long-lived persistent current states with an optimal operation time of only 10 ns.

This technology is fully compatible with SFQ-based control from on-chip digital-to-analog converters (DACs)

[26]. We showed our innovative preparation protocol where we used rapid adiabatic passage (RAP) to generate a single excitation in the FQ and swap it into the CQ. We also demonstrated a method for creating coherent superposition states, naturally preparing and reading out the CQ in the $X$ basis. These methods enable rapid qubit characterization and microwave-free spectroscopy.

Both the preparation and readout methods presented in this article have wide operating margins and are robust to changes in the CQ frequency allowing for the CQ frequency to be freely tuned without the need to recalibrate either the preparation or readout parameters. This simplifies the gate calibration process for superconducting qubits and broadens the interval over which that calibration remains stable.

Finally, when combined with two stages of QFP amplification, the signal generated by the readout was high enough to drive an analog-to-digital converter (ADC) [13]. We demonstrated that the QFP chain can achieve $10^{-5}$ error with anneal times of $<$2.5 ns for each QFP. The combination of these technologies with flux qubit readout enabled the complete transformation of quantum information to digital data in only 15 ns, which was an order of magnitude faster than conventional cQED readout techniques [1–4]. Once converted to digital data, the information could be used for logical operations without ever leaving the cold space. This offers a significant reduction in I/O signal counts and paves the way for superconducting quantum processors using scalable superconducting digital control and measurement electronics.

## ACKNOWLEDGMENTS

We are grateful to the technical and non-technical staff at Northrop Grumman Systems Corporation (NGSC) for supporting the advancement of our quantum computing architecture. We would like to thank John Fusco, John X. Przybysz, Chris Kirby, Jim Baumgardner, and Ofer Naaman for laying the groundwork at NGSC that allowed this work to flourish.

This work was supported in part by the United States Government.

### A. Author contributions

R. Abraham, F. Amet, P. Anderson, M. Arrigo, J. Arteaga, C. J. Ballard, C. Barker, T. Barnes, P. Bechman, R. Bhatt, K. Blaine, T. M. Borman, J. Botimer, G. R. Boyd, P. Bradley, A. D. Brandon, T. K. Bristol, A. Bulkos, R. M. Burnett, D. Burrowes, D. N. Cakan, N. Carniero, T. Chamberlin, D. Chen, M. Chilcote, B. G. Christensen, I. Christie, J. Clark, D. J. Clarke, J. M. Cochran, J. C. Collini, K. Connolly, G. Costa, D. M. Cowger, B. Dalfort, D. Davies, R. D. Dawson, S. Deitemeyer, N. DeNigris, F. Densmore, S. J. Di Giacomo, S. G. Diamond, R. DiCiro, S. M. Disseler, K. Dixon, J. Donnelly, E. Donohue, M. Downing, B. Eastin, N. Edwards, M. Ehsani, S. Ellis, R. Epstein, D. Ferguson, P. Fischer, T. Forzani, S. Friedensen, D. Gabriel, M. A. Gettelman, A. Gillam, G. Gilmore, J. Goode, E. Goodwin, M. Gottschalk, A. L. Graninger, T. Graves-Abe, J. Hackley, N. Hartman, B. Heacock, B. Heischmidt, P. Helms, R. Hinkey, B. Hong, S. T. Howard, D. Jensen, N. Johns, D. R. Johnson, J. Johnson, P. Kamenov, H. Kaplan, Z. Keane, S. Keebaugh, C. Kegerreis, K. Kelcourse-Oquendo, M. S. Khalil, D. Killeen, A. S. Knutson, T. Kohler, M. Kornecki, F. Koutsouli, A. Krick, K. L. Krycka, J. Kuan, D. Lad, J. R. Lane, N. J. Laurita, A. C. Lee, A. R. Lemmon, E. M. Leonard, L. M. D. Leonard, J. Leventis, M. P. Lilly, A. Lisewski, L. Llano, M. Longo, C. Lostoski, Z. Lou, M. G. Loving, D. al Ludwig, N. Luhrs, J. L. Lund, K. A. Maddock, R. J. Magyar, K. Mahmud, T. A. Manning, A. I. Marakov, R. Mays, A. D. McCreary, P. F. McLaughlin, J. R. Medford, E. Metz, A. L. Middleton, A. Miklich, R. Miller, J. Mlack, M. Mucci, N. Mungo, T. Murphy, R. E. Murray, O. Naaman, J. Nakamura, M. Noevere, S. Novikov, M. E. Nowakowski, C. Nunez, K. N. Ogg, B. Oshokoya, D. Paz, A. Pesetski, T. Pillsbury, C. Pinion, A. D. Pitcock, A. J. Przybysz, P. Quarterman, D. Queen, A. Ramanayaka, I. Ramos, S. L. Reed, D. Reitz, M. Rennie, S. Reza, E. Rhee, B. Richman, K. H. Rigdon, C. Rotella, M. Rudolph, E. Sadler, T. Safford, D. Saha, A. E. Saia, D. I. Santiago, R. Schwartz, A. Schwarzkopf, M. Scott-Jones, S. L. Sendelbach, S. J. Shapiro, A. N. Sharma, M. E. Sherwin, A. Sierakowski, J. N. Sills, R. Simha, N. Siwak, J. Smith, C. Snyder, H. Solomon, N. S. Sperlein, L. St. Marie, K. Stalpes, S. F. Steers, Z. Stegen, R. M. Stein, T. I. Stephenson, M. Stoutimore, J. D. Strand, J. A. Strong, E. Swain, S. Swami, A. Tamang, I. Thompson, A. B. Tokarchik, N. Tralshawala, D. Tran, S. Tyler, L. Upton, G. Van Dyke, J. Vannucci, K. Vempati, J. S. Vogel, R. Vyas, M. Wagner, P. Warner, N. Washington, B. Way, M. A. Wayne, D. Wehella-Gamage, M. E. Weippert, T. Weitzel, J. Wenner, L. White, S. Whitsitt, A. Wilkinson, R. Willey, I. Yandow, J. Yang, L. Yu, F. Yumiceva, V. Zheng, R. Zimmerman, C. Zinn, J. Ziskind

Each author's contribution(s) to the paper are listed below in alphabetical order according to the CRediT model.

Writing – original draft: JB, BGC, DJC, AJP, RMS

Writing – review & editing: JB, TC, BGC, DJC, GC, NH, MSK, EML, JRM, ALM, REM, AP, AJP, RMS, JDS

Conceptualization: CJB, JB, RMB, TC, BGC, DJC, DF, RH, MSK, JLL, KAM, RJM, TAM, AIM, JRM, RM, REM, ON, AP, AJP, RMS, MS, JDS, LU

Data Curation: FA, PA, JB, ADB, TC, BGC, JMC, GC, EG, MG, ALG, NH, BeH, RH, MSK, MK, AK, KLK, DL, NJL, EML, LMDL, JLL, KAM, AIM, JRM, ALM, JM, KNO, AJP, PQ, DQ, KHR, CR, ErS, SLS, AS, LS, RMS, JDS, GVD, PW, MAW, RW, JY, RZ, JZ

Formal Analysis: MA, CJB, CB, JB, GRB, ADB, RMB, TC, BGC, IC, DJC, KC, GC, DD, FD, RE, DF, MG, ALG, PH, RH, DRJ, HK, MSK, ASK, KLK, DL,

LMDL, NL, JLL, RJM, KM, AIM, JRM, ALM, AM, RM, REM, ON, KNO, DP, AJP, PQ, DQ, DR, BR, TS, DIS, MS-J, SLS, SJS, JNS, JS, KS, RMS, JDS, ABT, NT, BW, MEW, TW, SW, RZ

Investigation: RA, FA, PA, CJB, TMB, JB, RMB, TC, MC, BGC, JC, DJC, JMC, JCC, GC, DMC, BD, SD, ND, SGD, SMD, KD, JD, ED, MD, NE, SE, DF, SF, AG, EG, MG, ALG, JH, NH, BeH, BrH, RH, STH, DJ, NJ, HK, SK, CK, KK-O, MSK, ASK, TK, MK, AK, KLK, JK, DL, JRL, NJL, ACL, EML, LMDL, MPL, ML, CL, DlL, NL, JLL, KAM, RJM, TAM, AIM, ADM, JRM, EM, ALM, AM, RM, JM, NM, REM, ON, JN, SN, MEN, CN, KNO, TP, CP, AJP, PQ, DQ, SLR, DR, MR, KHR, CR, MaR, ErS, TS, RoS, AnS, MS-J, SLS, ANS, JNS, RS, NS, JS, CS, HS, LS, KS, SFS, ZS, RMS, TIS, MS, JDS, IT, ABT, NT, ST, LU, JV, JSV, MW, PW, NW, MAW, DWG, JW, AW, IY, LY, FY, VZ, RZ, CZ

Methodology: PA, MA, JA, CJB, CB, TB, RB, KB, JB, GRB, PaB, ADB, TKB, AB, RMB, DB, DNC, NC, TC, DC, BGC, IC, JC, DJC, GC, DD, FD, RD, SMD, KD, JD, ME, SE, RE, DF, PF, DG, GG, JG, MG, ALG, BeH, BrH, PH, RH, BH, NJ, DRJ, HK, CK, KK-O, MSK, DK, ASK, TK, FK, AK, KLK, DL, JRL, NJL, ACL, EML, LMDL, MPL, LL, CL, ZL, NL, JLL, KAM, RJM, KM, TAM, AIM, ADM, PFM, JRM, ALM, AM, RM, JM, MM, REM, ON, JN, SN, MEN, KNO, BO, DP, CP, ADP, AJP, PQ, DQ, IR, DR, SR, BR, KHR, CR, MaR, ErS, TS, DS, DIS, AnS, MS-J, SLS, SJS, ANS, JNS, JS, HS, NSS, KS, SFS, ZS, RMS, MS, JDS, JAS, ES, SS, AT, IT, ABT, NT, DT, ST, LU, KV, JSV, MW, PW, NW, MAW, DWG, MEW, TW, JW, LW, SW, AW, RZ, CZ

Project Administration: CJB, JB, RMB, TC, DC, BGC, JC, BE, MG, ALG, BeH, RH, ZK, MSK, AK, NJL, ACL, EML, LMDL, MPL, JLL, AIM, JRM, REM, ON, AP, CP, AJP, PQ, DQ, KHR, ErS, AES, SLS, AS, JS, SFS, RMS, MS, JDS, LU

Resources: RA, PB, TMB, JB, DNC, SJDG, ED, MAG, ALG, TGA, JH, RH, JJ, SK, MSK, ACL, MPL, AL, MGL, KAM, ADM, PFM, JRM, JM, NM, REM, MN, KNO, TP, CP, AJP, PQ, DQ, MR, KHR, CR, ErS, AES, AS, RS, NS, CS, RMS, JDS, LU, JV, JSV, PW, MAW, TW, LW, VZ, RZ

Software: FA, PA, CJB, CB, JB, GRB, ADB, RMB, TC, DC, IC, DJC, KC, GC, BD, DD, SD, ND, FD, SGD, SMD, JD, MD, BE, NE, RE, DF, SF, EG, MG, ALG, NH, BeH, PH, RH, STH, DJ, DRJ, HK, MSK, ASK, TK, KLK, JK, DL, JRL, NJL, EML, JL, CL, DlL, NL, JLL, KAM, RJM, KM, TAM, AIM, ADM, PFM, JRM, EM, ALM, AM, RM, JM, TM, REM, ON, JN, SN, MEN, CN, KNO, DP, AJP, DQ, DR, BR, KHR, MaR, TS, DIS, RoS, AnS, MS-J, SLS, SJS, AS, JNS, JS, LS, KS, ZS, RMS, TIS, MS, JDS, JAS, IT, ABT, NT, LU, GVD, JV, JSV, BW, MAW, MEW, TW, JW, LW, SW, RW, FY, RZ, CZ

Supervision: PA, CJB, JB, ADB, RMB, TC, BGC, IC, JC, DJC, JMC, GC, BD, BE, DF, ALG, NH, BeH, PH, RH, HK, CK, MSK, KLK, NJL, ACL, EML, LMDL, MPL, AL, ML, CL, NL, JLL, KAM, AIM, ADM, JRM, AM, RM, JM, REM, ON, MN, KNO, AP, CP, AJP, PQ, DQ, MR, KHR, ErS, TS, DS, AES, MS-J, SLS, AS, JS, HS, LS, KS, SFS, RMS, MS, JDS, IT, ST, LU, JV, JSV, MW, PW, MAW, MEW, JW, LW, RZ

Validation: CJB, CB, JB, ADB, RMB, TC, BGC, IC, DJC, GC, DF, MG, ALG, RH, HK, MSK, KLK, EML, LMDL, JLL, KAM, RJM, AIM, JRM, ALM, RM, DP, AJP, DR, TS, SLS, AS, KS, RMS, JDS, ABT, NT, RZ

Visualization: JB, TC, BGC, DJC, GC, RE, NH, MSK, KLK, EML, RJM, JRM, ALM, REM, AJP, RMS

Funding Acquisition: CJB, JB, BGC, DF, ACL, EML, MPL, JRM, AP, AJP, MES, MS, JDS

### B. Competing interests

Authors declare that they have no competing interests.

### C. Data, code, and materials availability

Data and code used to generate figures in this paper are available in a public repository. Although they are approved for public release, these data and code remain the property of Northrop Grumman Systems Corporation.

## SUPPLEMENTARY MATERIALS

### Appendix A: Effective Circuit Hamiltonian

The theoretical model of the computational qubit (CQ) and flux qubit (FQ) combined system was a numerical model with a Hamiltonian derived from the circuit schematic shown in Fig. S1A. The degrees of freedom in the model were the node charges, $\hat{n}_i$, of the three nodes in the circuit and the phase drops, $\hat{\phi}_i$, between nodes. The Hamiltonian of the circuit was

$$H = \frac{\hbar}{2Z_0}\,\hat{\vec{n}} \cdot C^{-1} \cdot \hat{\vec{n}} + \frac{\hbar Z_0}{2}\left(\hat{\vec{\phi}} \cdot M^T + 2\pi\vec{\alpha}\right) \cdot \Gamma \cdot \left(M \cdot \hat{\vec{\phi}} + 2\pi\vec{\alpha}\right) + H_{J_i} - \sum_j \frac{\hbar}{2e}\, I_{c_j} \cos\!\left(e_j \cdot (M \cdot \hat{\vec{\phi}} + 2\pi\vec{\alpha})\right) \quad \text{(S1)}$$

where $\hbar = 1.055 \times 10^{-34}$ Js is Planck's constant over $2\pi$, $Z_0 = 25.8$ k$\Omega$ is the impedance quantum, and $e = 1.602 \times 10^{-19}$ C is the electron charge. $C^{-1}$ is the inverse of the capacitance matrix, $\Gamma$ is the inverse inductance matrix, and M is the incidence matrix [27]. The last term is a sum over all the Josephson junction elements where $I_{c_j}$ is the critical current of each one and $e_j$ is a unit vector in the direction of the element. Finally, $H_{J_i}$ is the contribution to the Hamiltonian from the two large junctions in the FQ between nodes 0 and 1. It is an effective series junction model

$$H_{J_i} = -\frac{\hbar}{16 Z_0 C_J}\left(1 - \frac{1}{N}\right)^2 (y - \sqrt{y}) \quad \text{(S2)}$$

$$y e^{1/\sqrt{y}} = 16N^2 \left(\frac{Z_0}{Z}\right)^2 \left(1 - \frac{1}{N}\right)^{-2} \cos\!\left(\frac{\phi}{N} - 2\pi\Delta\right) \quad \text{(S3)}$$

where $N = 2$ is the number of junctions in series, $C_J$ is the capacitance of each junction, $Z = \sqrt{\frac{2\pi I_c}{\Phi_0 C_J}}$ is the impedance of each junction, $\phi$ is the phase drop across the effective circuit element, and $\Delta$ is the flux applied to the element in magnetic flux quanta, $\Phi_0 = 2.068 \times 10^{-15}$ Wb. We performed a 4th order expansion of Eq. S3 in the small parameter

$$x = 1 - \cos\!\left(\frac{\phi}{N} - 2\pi\Delta\right) \quad \text{(S4)}$$

before promoting the phase drop, $\phi$, across the effective circuit element to a phase operator, $\hat{\varphi}$. This effective inductor model behaved as a linear inductance and improved the efficiency of the model by reducing the node count.

Fig. S1B shows probe tone spectroscopy of a device presented in this article. The energy spectrum in Fig. S1B is overlaid with a Hamiltonian simulation for the same flux values, and was inclusive of junction asymmetry of 1%, which created the asymmetric nature of the spectra in this example device. The circuit Hamiltonian had up to 31, 9, and 9 excitations on the 1st, 2nd and 3rd degrees of freedom, respectively, giving a Hilbert space dimension of 2511. We performed a time-dependent simulation of this model with ramped $\alpha$ and $\Delta$ flux control to create the dashed traces in Fig. S1B.

### Appendix B: Flux Qubit Readout of Higher Excited States

We could discriminate the flux qubit MRT regions in a measurement of the probability of measuring the flux qubit in a given well versus the applied $\Delta$ flux (Fig. S2A). At $\Delta = 0$, the ground states in the two wells were degenerate, and the probability of reading the right circulating current, $P_{|R\rangle}$ , transitioned through 0.5. At $\Delta$ <0, the qubit ground state projected into the left ($|L\rangle$) well, while at $\Delta$ >0 it projected into the right ($|R\rangle$) well. In the 1st MRT window, the qubit $|0\rangle$ and $|1\rangle$ states projected into different wells. In the 2nd MRT window, the $|0\rangle$ and $|1\rangle$ states both projected into the same well while the $|2\rangle$ state ended up as the lowest energy state in the other well, and so on for higher MRTs.

Fig. S2A shows the result of an experiment where we intentionally heated the cryostat such that the sample was at an elevated temperature T=130 mK. We prepared the system in a thermal equilibrium state and measured the flux qubit $\Delta$ transition. Instead of a smooth S-curve, we observed relatively flat "shoulders" in each of the MRT regions, each of which represented measurements of the thermal populations of the computational qubit excited states (i.e. $|1\rangle$, $|2\rangle$, and $|3\rangle$). The dashed line in Fig. S2A is the result of a unitary time-dependent simulation of the Hamiltonian shown in Equation S1 where the system began in a thermal equilibrium state with T=130 mK. The slightly asymmetric transition, where the 1st MRT shoulder is more rounded at positive FQ $\Delta$ flux, was due to junction asymmetry and was captured in the simulation.

Fig. S2B shows the result of an experiment where we used RAP prep to prepare the CQ in the $|1\rangle$ and delayed before measuring to observe the $T_1$ of the CQ. By choosing different readout points in each MRT, we sampled the populations of the 4 lowest energy states during the decay and observed a $T_1 = 50$ $\mu$s for the CQ. Fig. S2C shows an experiment where we adjusted the offset and amplitude of the RAP prep pulse such that we prepared the computational qubit in the $|2\rangle$ state and observed its decay. Note the initial rise in the $|1\rangle$ population as the state decayed one photon at a time from $|2\rangle$ to $|1\rangle$ and eventually to $|0\rangle$. The dots are the data and the solid lines are a fit to the Lindblad master equation [28, 29], illustrating our ability to accurately measure the dynamics of higher order states.

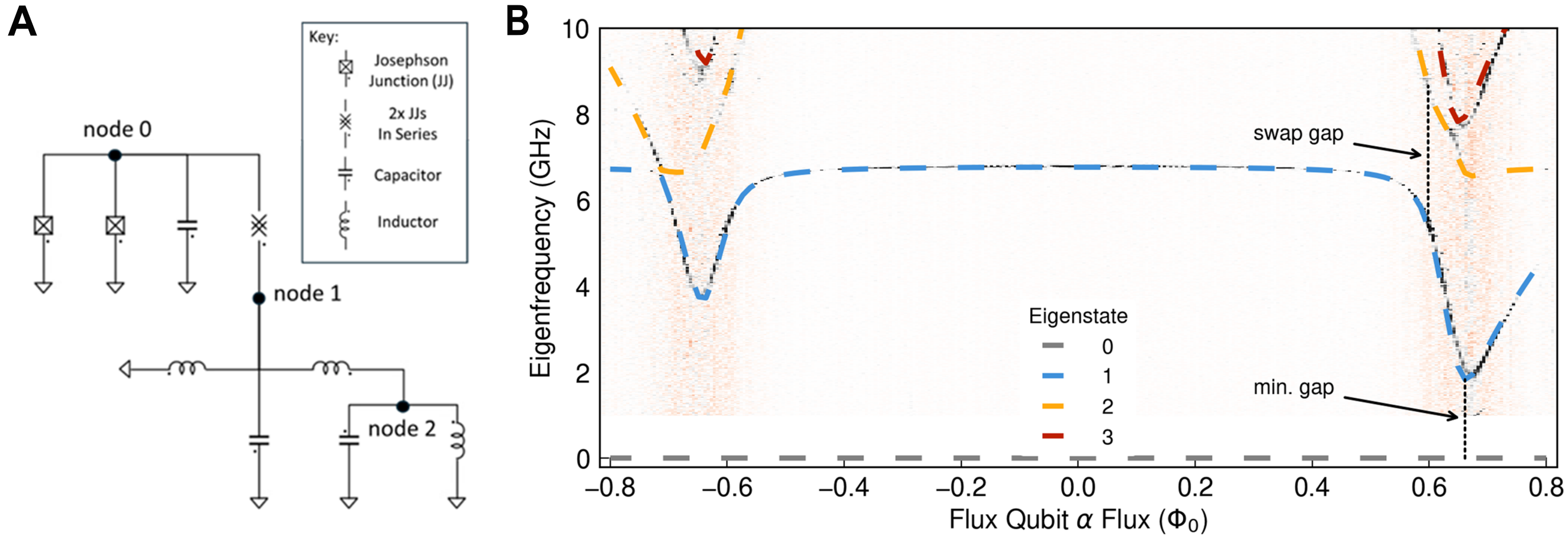

FIG. S1. Circuit schematic for Hamiltonian simulation. (A) Schematic for the computational qubit (CQ) and flux qubit (FQ) combined. The symbol with overlapping Xs is an effective circuit element that represents two Josephson junctions in series. (B) Probe tone spectroscopy of the device with simulation overlaid. The dashed lines are the energy spectrum determined from the circuit Hamiltonian.

### Appendix C: Photon Loss During Readout

The dominant source of SPAM infidelity, for the devices discussed in the main text, was photon loss. There were two primary sources of photon loss: dielectric loss and swapping the qubit state into strongly-coupled two-level systems (TLS) during the anneal sequence. First, while the computational qubit was designed to have long coherence times, the readout flux qubit was instead optimized to have a fast readout speed with minimal impact to the overall qubit footprint. To accomplish this objective, the flux qubits had a low shunt capacitance and a small footprint, both of which were at odds with typical long coherence qubit designs. As such, the typical quality factor of our flux qubits was around 200k, which set a median infidelity from dielectric loss of approximately $10^{-3}$.

The more concerning loss, though, was from strongly-coupled TLS, where the qubit state swapped into the TLS as the flux qubit frequency was swept through a TLS during the readout anneal sequence. For example, with our flux qubits we expected approximately $10^{-3}$ infidelity from a TLS with a 10 MHz coupling strength, i.e., a single TLS with 10 MHz coupling strength imposed as much infidelity loss as the entire bath of weakly-coupled TLS. The fields inside the junction reached a few kV/m, which allowed for coupling strengths as high as 300 MHz. In this interface, TLS with coupling strengths greater than 10 MHz are not uncommon [30]. TLS in the junction were relatively rare in frequency space; however, the flux qubit underwent a large frequency excursion during readout (on order of 4 GHz) after swapping with the computational qubit.

Studies of our junctions have estimated the material constant to be $\sigma = 0.22 \pm 0.02$ GHz$^{-1}$ $\mu$m$^{-2}$, which is in line with published estimates [30, 31]. For perspective, this material constant gives about 0.16 TLS/GHz with a coupling strength $g \geq 10$ MHz; therefore, half of the qubits would interact with a TLS with appreciable coupling strengths during readout. As the infidelity loss from a TLS is expected to follow the Landau-Zener transition probability $\left(P_e \approx 1 - \exp\left(-\frac{\pi^2 g^2}{\nu}\right) \approx \frac{\pi^2 g^2}{\nu}\right)$, where g is the coupling strength, and $\nu$ is the energy velocity, strongly-coupled TLS should scale quadratically with the coupling strength. As the coupling strength of these junction-based TLS can exceed 100 MHz, the interactions of which are expected to cause an infidelity of 0.1. This crude analysis explains the devices with low SPAM error; however, there were a plethora of TLS below 10 MHz that provided photon loss.

A representative plot that shows these effects is shown in Fig. S3 where we plot photon decay rate versus flux qubit $\alpha$ flux. When interpreting this plot, it is important to reference Fig. 1D in the main text to understand how the mode frequency varied with $\alpha$ bias. At low $\alpha$ (0 – 0.5 $\Phi_0$), the mode was predominantly in the computational qubit and underwent small frequency excursions from the weak hybridization with the flux qubit mode. As only a handful of MHz were swept during the first 0.5 $\Phi_0$, weakly coupled TLS would appear as apparent wide peaks in the decay rate during this region. Around 0.5 $\Phi_0$ the computational qubit state began to adiabatically transition to the flux qubit mode, which led to an increase in the decay rate as the flux qubit mode participation increased. Around 0.7 $\Phi_0$, the barrier in the double-well potential of the flux qubit began to separate the circulating current states, providing protection from energy relaxation. The handful of peaks in the region where the mode has swapped to the flux qubit (0.5 – 0.7 $\Phi_0$) were due to interactions with strongly-coupled

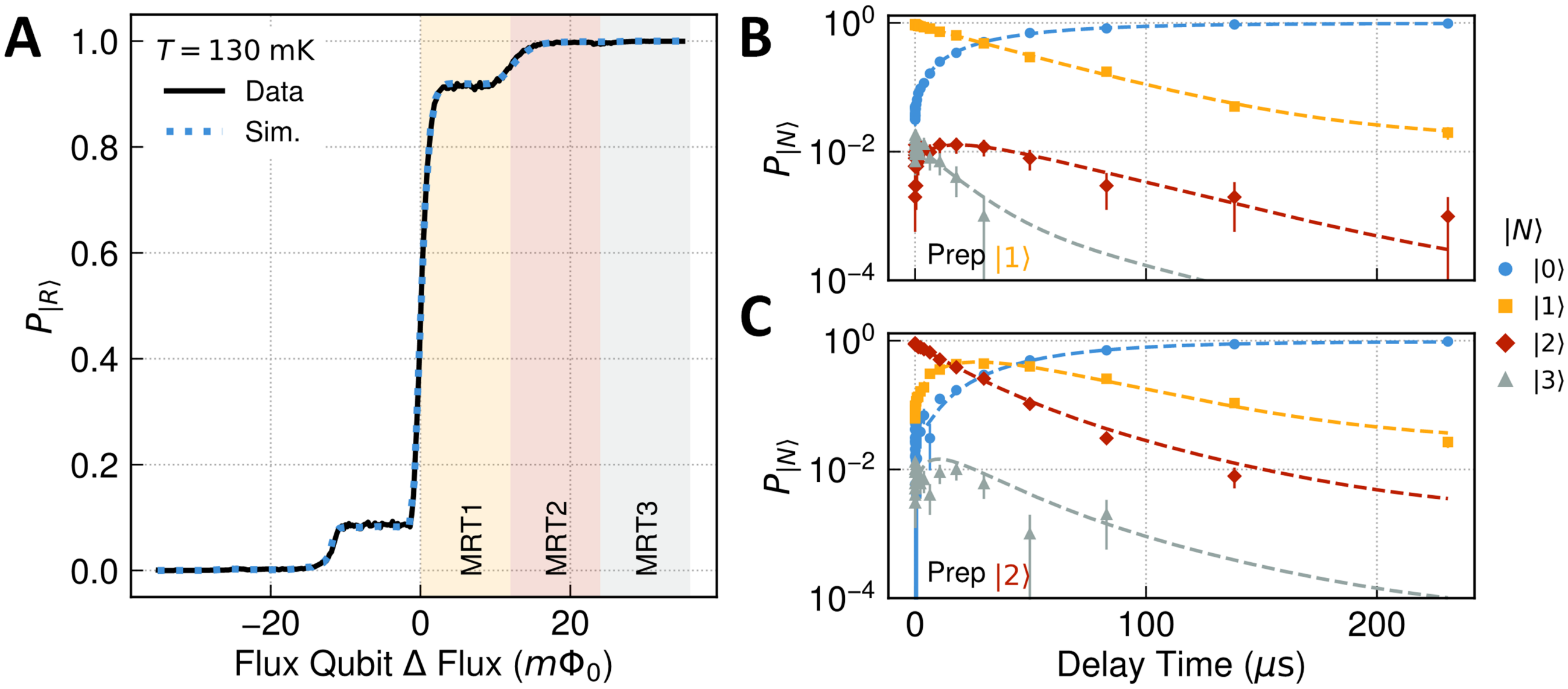

FIG. S2. Flux Qubit Readout of Higher Excited States. (A) The probability of the flux qubit (FQ) projecting into the right well as a function of the Δ flux. As Δ flux was adjusted the double-well potential was tilted, changing the state ordering between wells. At zero Δ flux the two wells were degenerate. We found that increasing the Δ flux to higher MRT windows would tune which computational qubit (CQ) state populations were detected. Here the device was intentionally heated to 130 mK to elevate the populations of the CQ excited states and thus make the MRT regions more visible. The dashed line is not a fit but rather the result of a time-dependent simulation of the circuit Hamiltonian model derived from fitting the spectra in Fig. S1B. (B) Multi-state time dynamics of a CQ prepared in $|1\rangle$ using RAP prep. (C) Multi-state time dynamics of a CQ prepared in the $|2\rangle$ state using RAP prep. This CQ had $T_1 = 50$ $\mu$s and high $|2\rangle$ preparation fidelity. For B and C, the data was acquired by repeatedly measuring in each of the first three MRTs and deconvolving the combined measurements. The data was fit using the Lindblad master equation [26–28].

TLS.

Improving the readout decay constant, $\tau_D$, will require improvements to the flux noise and, more importantly, improvements to the junction barrier quality. Recent work [32] has shown viable paths to reducing the number density of strongly-coupled TLS. Continued research into barrier improvements is critical for fast adiabatic readout technologies.

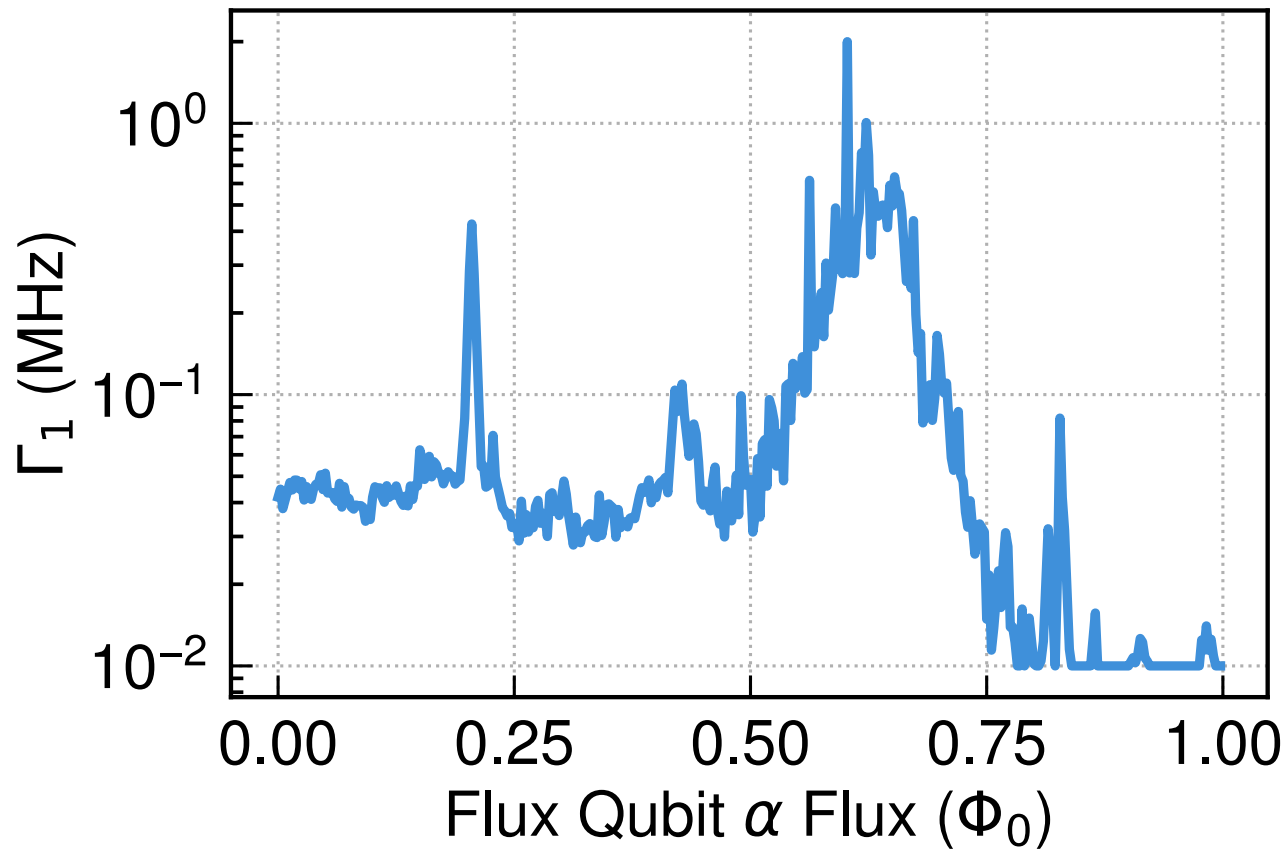

FIG. S3. TLS Loss During Flux Qubit Annealing. The qubit mode decay rate (Γ) as a function of the flux qubit $\alpha$ flux. The peak near 0.2 $\Phi_0$ was caused by a weakly-coupled TLS in the computational qubit capacitor. As the mode transitioned from the computational qubit to the flux qubit (around 0.5 $\Phi_0$), the background decay rate increased due to the lower quality factor of the flux qubit. The peaks in the flux qubit region of the curve correspond to interactions with strongly coupled TLS (coupling strengths exceeding 5 MHz).

[1] M. AbuGhanem. Ibm quantum computers: evolution, performance, and future directions. *J. Supercomput.*, 81:687, 2025.

[2] D. T. McClure, H. Paik, L. S. Bishop, M. Steffen, J. M. Chow, and J. M. Gambetta. Rapid driven reset of a qubit readout resonator. *Phys. Rev. Appl.*, 5:011001, 2016.

[3] T. White, A. Opremcak, G. Sterling, A. Korotkov, D. Sank, R. Acharya, M. Ansmann, F. Arute, K. Arya, J. C. Bardin, A. Bengtsson, A. Bourassa, J. Bovaird, L. Brill, B. B. Buckley, D. A. Buell, T. Burger, B. Burkett, N. Bushnell, Z. Chen, B. Chiaro, J. Cogan, R. Collins, A. L. Crook, B. Curtin, S. Demura, A. Dunsworth, C. Erickson, R. Fatemi, L. Flores Burgos, E. Forati, B. Foxen, W. Giang, M. Giustina, A. Grajales Dau, M. C. Hamilton, S. D. Harrington, J. Hilton, M. Hoffmann, S. Hong, T. Huang, A. Huff, J. Iveland, E. Jeffrey, M. Kieferová, S. Kim, P. V. Klimov, F. Kostritsa, J. M. Kreikebaum, D. Landhuis, P. Laptev, L. Laws, K. Lee, B. J. Lester, A. Lill, W. Liu, A. Locharla, E. Lucero, T. McCourt, M. McEwen, X. Mi, K. C. Miao, S. Montazeri, A. Morvan, M. Neeley, C. Neill, A. Nersisyan, J. H. Ng, A. Nguyen, M. Nguyen, R. Potter, C. Quintana, P. Roushan, K. Sankaragomathi, K. J. Satzinger, C. Schuster, M. J. Shearn, A. Shorter, V. Shvarts, J. Skruzny, W. Clarke Smith, M. Szalay, A. Torres, B. W. K. Woo, Z. Jamie Yao, P. Yeh, J. Yoo, G. Young, N. Zhu, N. Zobrist, Y. Chen, A. Megrant, J. Kelly, and O. Naaman. Readout of a quantum processor with high dynamic range Josephson parametric amplifiers. *Appl. Phys. Lett.*, 122:014001, 2023.

[4] T. Walter, P. Kurpiers, S. Gasparinetti, P. Magnard, A. Potočnik, Y. Salathé, M. Pechal, M. Mondal, M. Oppliger, C. Eichler, and A. Wallraff. Rapid high-fidelity single-shot dispersive readout of superconducting qubits. *Phys. Rev. Appl.*, 7:054020, 2017.

[5] P. A. Spring, L. Milanovic, Y. Sunada, S. Wang, A. F. van Loo, S. Tamate, and Y. Nakamura. Fast multiplexed superconducting-qubit readout with intrinsic Purcell filtering using a multiconductor transmission line. *PRX Quantum*, 6:020345, 2025.

[6] R. Acharya, I. Aleiner, R. Allen, T. I. Andersen, M. Ansmann, F. Arute, K. Arya, A. Asfaw, J. Atalaya, R. Babbush, D. Bacon, J. C. Bardin, J. Basso, A. Bengtsson, S. Boixo, G. Bortoli, A. Bourassa, J. Bovaird, L. Brill, M. Broughton, B. B. Buckley, D. A. Buell, T. Burger, B. Burkett, N. Bushnell, Y. Chen, Z. Chen, B. Chiaro, J. Cogan, R. Collins, P. Conner, W. Courtney, A. L. Crook, B. Curtin, D. M. Debroy, A. Del Toro Barba, S. Demura, A. Dunsworth, D. Eppens, C. Erickson, L. Faoro, E. Farhi, R. Fatemi, L. Flores Burgos, E. Forati, A. G. Fowler, B. Foxen, W. Giang, C. Gidney, D. Gilboa, M. Giustina, A. Grajales Dau, J. A. Gross, S. Habegger, M. C. Hamilton, M. P. Harrigan, S. D. Harrington, O. Higgott, J. Hilton, M. Hoffmann, S. Hong, T. Huang, A. Huff, W. J. Huggins, L. B. Ioffe, S. V. Isakov, J. Iveland, E. Jeffrey, Z. Jiang, C. Jones, P. Juhas, D. Kafri, K. Kechedzhi, J. Kelly, T. Khattar, M. Khezri, M. Kieferová, S. Kim, A. Kitaev, P. V. Klimov, A. R. Klots, A. N. Korotkov, F. Kostritsa, J. M. Kreikebaum, D. Landhuis, P. Laptev, K.-M. Lau, L. Laws, J. Lee, K. Lee, B. J. Lester, A. Lill, W. Liu, A. Locharla, E. Lucero, F. D. Malone, J. Marshall, O. Martin, J. R. McClean, T. McCourt, M. McEwen, A. Megrant, B. Meurer Costa, X. Mi, K. C. Miao, M. Mohseni, S. Montazeri, A. Morvan, E. Mount, W. Mruczkiewicz, O. Naaman, M. Neeley, C. Neill, A. Nersisyan, H. Neven, M. Newman, J. H. Ng, A. Nguyen, M. Nguyen, M. Y. Niu, T. E. O'Brien, A. Opremcak, J. Platt, A. Petukhov, R. Potter, L. P. Pryadko, C. Quintana, P. Roushan, N. C. Rubin, N. Saei, D. Sank, K. Sankaragomathi, K. J. Satzinger, H. F. Schurkus, C. Schuster, M. J. Shearn, A. Shorter, V. Shvarts, J. Skruzny, V. Smelyanskiy, W. Clarke Smith, G. Sterling, D. Strain, M. Szalay, A. Torres, G. Vidal, B. Villalonga, C. Vollgraff Heidweiller, T. White, C. Xing, Z. Jamie Yao, P. Yeh, J. Yoo, G. Young, A. Zalcman, Y. Zhang, and N. Zhu. Suppressing quantum errors by scaling a surface code logical qubit. *Nature*, 614:676–681, 2023.

[7] T. C. White, J. Y. Mutus, I.-C. Hoi, R. Barends, B. Campbell, Y. Chen, Z. Chen, B. Chiaro, A. Dunsworth, E. Jeffrey, J. Kelly, A. Megrant, C. Neill, P. J. J. O'Malley, P. Roushan, D. Sank, A. Vainsencher, J. Wenner, S. Chaudhuri, J. Gao, and J. M. Martinis. Traveling wave parametric amplifier with Josephson junctions using minimal resonator phase matching. *Appl. Phys. Lett.*, 106:242601, 2015.

[8] L. Planat, R. Dassonneville, J. Puertas Martínez, F. Foroughi, O. Buisson, W. Hasch-Guichard, C. Naud, R. Vijay, K. Murch, and N. Roch. Understanding the saturation power of Josephson parametric amplifiers made from SQUID arrays. *Phys. Rev. Appl.*, 11:034014, 2019.

[9] J. Yoo, Z. Chen, F. Arute, S. Montazeri, M. Szalay, C. Erickson, E. Jeffrey, R. Fatemi, M. Giustina, M. Ansmann, E. Lucero, J. Kelly, and J. C. Bardin. Design and characterization of a $<$ 4-mw/qubit 28-nm cryo-CMOS integrated circuit for full control of a superconducting quantum processor unit cell. *IEEE J. Solid-State Circuits*, 58:3044–3060, 2023.

[10] C. H. Liu, A. Ballard, D. Olaya, D. R. Schmidt, J. Biesecker, T. Lucas, J. Ullom, S. Patel, O. Rafferty, A. Opremcak, K. Dodge, V. Iaia, T. McBroom, J. L. DuBois, P. F. Hopkins, S. P. Benz, B. L. T. Plourde, and R. McDermott. Single flux quantum-based digital control of superconducting qubits in a multichip module. *PRX Quantum*, 4:030310, 2023.

[11] C. Jordan, J. Bernhardt, J. Rahamim, A. Kirichenko, K. Bharadwaj, L. Fry-Bouriaux, A. Somoroff, K. Porsch, K.-T. Tsai, J. Walter, A. Weis, M.-J. Yu, M. Renzullo, J. Javelle, C. Checkley, O. Mukhanov, D. Yohannes, I. Vernik, and S.-J. Han. A quantum computer controlled by superconducting digital electronics at millikelvin temperature. *Nat. Electron.*, 9:287–294, 2026.

[12] J. A. Grover, J. I. Basham, A. Marakov, S. M. Disseler, R. T. Hinkey, M. Khalil, Z. A. Stegen, T. Chamberlin, W. DeGottardi, D. J. Clarke, J. R. Medford, J. D. Strand, M. J. A. Stoutimore, S. Novikov, D. G. Ferguson, D. Lidar, K. M. Zick, and A. J. Przybysz. Fast, lifetime-preserving readout for high-coherence quantum annealers. *PRX Quantum*, 1:020314, 2020.

[13] C. Ballard, A. Miklich, M. Stoutimore, R. Miller, J. Strand, and K. Pleim. Flux switch system. U.S. Patent 11 486 910 B1, 2022.

[14] L. F. Wei, J. R. Johansson, L. X. Cen, S. Ashhab, and F. Nori. Controllable coherent population transfers in superconducting qubits for quantum computing. *Phys. Rev. Lett.*, 100:113601, 2008.

[15] M. Steffen, F. Brito, D. DiVincenzo, M. Farinelli, G. Keefe, M. Ketchen, S. Kumar, F. Milliken, M. B. Rothwell, and J. Rozen. Quantum information storage using tunable flux qubits. *J. Phys.: Condens. Matter*, 22:053201, 2010.

[16] L. Landau. On the theory of transfer of energy at collisions II. *Phys. Z. Sowjetunion*, 2:46–51, 1932.

[17] C. Zener. Non-adiabatic crossing of energy levels. *Proc. R. Soc. Lond. A*, 137:696–702, 1932.

[18] C. Quintana. *Superconducting flux qubits for high-connectivity quantum annealing without lossy dielectrics.* PhD thesis, University of California, Santa Barbara, 2017.

[19] N. Takeuchi, D. Ozawa, Y. Yamanashi, and N. Yoshikawa. An adiabatic quantum-flux-parametron as an ultra-low-power logic device. *Supercond. Sci. Technol.*, 26:035010, 2013.

[20] Y. Yamanashi, H. Matsushima, T. Ortlepp, and N. Yoshikawa. Evaluation of current sensitivity of quantum flux parametron. *Supercond. Sci. Technol.*, 30:084004, 2017.

[21] R. Harris, M. W. Johnson, T. Lanting, A. J. Berkley, J. Johansson, P. Bunyk, E. Ladizinsky, N. Ladizinsky, T. Oh, I. Perminov, C. Rich, M. C. Thom, E. Tolkacheva, C. J. S. Truncik, S. Uchaikin, J. Wang, B. Wilson, and G. Rose. Experimental investigation of an eight-qubit unit cell in a superconducting optimization processor. *Phys. Rev. B*, 82:024511, 2010.

[22] S. Miyajima, M. Tanaka, H. Akaike, Y. Yamanashi, T. Ortlepp, and N. Yoshikawa. Current sensitivity enhancement of a quasi-one-junction SQUID comparator with an input transformer. *J. Low Temp. Phys.*, 176:465–470, 2014.

[23] R. Harris, M. W. Johnson, S. Han, A. J. Berkley, J. Johansson, P. Bunyk, E. Ladizinsky, S. Govorkov, M. C. Thom, S. Uchaikin, B. Bumble, A. Fung, A. Kaul, A. Kleinsasser, M. H. S. Amin, and D. V. Averin. Probing noise in flux qubits via macroscopic resonant tunneling. *Phys. Rev. Lett.*, 101:117003, 2008.

[24] F. Yan, S. Gustavsson, A. Kamal, J. Birenbaum, A. P. Sears, D. Hover, T. J. Gudmundsen, D. Rosenberg, G. Samach, S. Weber, J. L. Yoder, T. P. Orlando, J. Clarke, A. J. Kerman, and W. D. Oliver. The flux qubit revisited to enhance coherence and reproducibility. *Nat. Commun.*, 7:12964, 2016.

[25] J. A. Strong, V. V. Talanov, M. E. Nielsen, A. C. Brownfield, N. Bailey, Q. P. Herr, and A. Y. Herr. A resonant metamaterial clock distribution network for superconducting logic. *Nat. Electron.*, 5(3):171–177, 2022.

[26] O. Naaman, D. Miller, and R. Burnett. Superconducting bi-directional current driver. U.S. Patent 10 122 351, 2018.

[27] G. Burkard, R. H. Koch, and D. P. DiVincenzo. Multilevel quantum description of decoherence in superconducting qubits. *Phys. Rev. B*, 69:064503, 2004.

[28] G. Lindblad. On the generators of quantum dynamical semigroups. *Commun. Math. Phys.*, 48:119–130, 1976.

[29] V. Gorini, A. Kossakowski, and E. C. G. Sudarshan. Completely positive dynamical semigroups of $N$-level systems. *J. Math. Phys.*, 17:821–825, 1976.

[30] J. M. Martinis, K. B. Cooper, R. McDermott, M. Steffen, M. Ansmann, K. D. Osborn, K. Cicak, S. Oh, D. P. Pappas, R. W. Simmonds, and C. C. Yu. Decoherence in Josephson qubits from dielectric loss. *Phys. Rev. Lett.*, 95:210503, 2005.

[31] D. C. Zanuz, Q. Ficheux, L. Michaud, A. Orekhov, K. Hanke, A. Flasby, M. B. Panah, G. J. Norris, M. Kerschbaum, A. Remm, F. Swiadek, C. Hellings, S. Lazar, C. Scarato, N. Lacroix, S. Krinner, C. Eichler, A. Wallraff, and J.-C. Besse. Mitigating losses of superconducting qubits strongly coupled to defect modes. *Phys. Rev. Appl.*, 23:044054, 2025.

[32] O. F. Wolff, H. Mantry, R. Raja, W. Peng, K. Singirikonda, S. Lee, S. Sudhaman, R. Goncalves, P. Y. Huang, A. Kou, and W. Pfaff. Structural control of two-level defect density revealed by high-throughput correlative measurements of Josephson junctions. arXiv:2602.11469, 2026.